\documentclass[conference]{IEEEtran}
\IEEEoverridecommandlockouts
\usepackage[top=0.75in, bottom=0.75in, left=0.625in, right=0.625in]{geometry} % Set margins HERE

\usepackage{cite}
\usepackage{amsmath,amssymb,amsfonts}
\usepackage{algorithm}
\usepackage {algpseudocode}
\usepackage{graphicx}
\usepackage{caption}
\usepackage{textcomp}
\usepackage{xcolor}
 \usepackage{float} 
 \usepackage{diagbox}
 \usepackage{subfig}
 \usepackage{url}
 \usepackage{hyperref}
 \usepackage{mathrsfs}
 \usepackage{mathtools}
 \usepackage{amsmath}
\usepackage{bm}
\newcommand{\card}[1]{\lvert #1 \rvert}
\usepackage{mleftright}
  \usepackage{bm}
  \usepackage{booktabs}
\def\BibTeX{{\rm B\kern-.05em{\sc i\kern-.025em b}\kern-.08em
    T\kern-.1667em\lower.7ex\hbox{E}\kern-.125emX}}

\makeatletter

\newcounter{phase}[algorithm]
\newlength{\phaserulewidth}
\newcommand{\phaseTitle}{Stage} % <-- Enter default "phase name" here
\newcommand{\setphaserulewidth}{\setlength{\phaserulewidth}}

\newcommand{\topPhase}[1]{%
  \vspace{2.0ex}
  \Statex\strut\refstepcounter{phase}\item[\textbf{\phaseTitle~\thephase:~}\textit{~#1}] % phase "caption"
   \vspace{-3.0ex}\Statex\leavevmode\llap{\rule{\dimexpr\labelwidth+\labelsep}{\phaserulewidth}}\rule{\linewidth}{\phaserulewidth}}

\newcommand{\phase}[1]{%
  \vspace{-1.25ex}
  \Statex\leavevmode\llap{\rule{\dimexpr\labelwidth+\labelsep}{\phaserulewidth}}\rule{\linewidth}{\phaserulewidth}
  \Statex\strut\refstepcounter{phase}\item[\textbf{\phaseTitle~\thephase:~}\textit{~#1}] % phase "caption"
  \vspace{-1.25ex}\Statex\leavevmode\llap{\rule{\dimexpr\labelwidth+\labelsep}{\phaserulewidth}}\rule{\linewidth}{\phaserulewidth}}

\makeatother

\setphaserulewidth{.7pt}

\usepackage[utf8]{inputenc}
\usepackage[T1]{fontenc}
    
\begin{document}

\title{Real-Time Line Parameter Estimation Method for Multi-Phase Unbalanced Distribution Networks\\
\thanks{The work is partially funded by the Natural Sciences and Engineering Research Council (NSERC) Discovery Grant, NSERC RGPIN-2022-03236.}
}

\author{ Sakirat Wolly, \textit{Student Member}, IEEE and Xiaozhe Wang, \textit{Senior Member}, IEEE \\ Department of Electrical and Computer Engineering, McGill University, Montreal, QC H3A 0E9, Canada \\ sakirat.wolly@mail.mcgill.ca, xiaozhe.wang2@mcgill.ca}

\maketitle
    
\begin{abstract}
%\color{red} update abstract giving all my modifications \color{black}
An accurate distribution network model is crucial for monitoring, state estimation and energy management. However, existing data-driven methods often struggle with scalability or impose a heavy computational burden on large distribution networks. In this paper, leveraging natural load dynamics, we propose a two-stage line estimation method for multiphase unbalanced distribution networks. Simulation results using real-life load and PV data show that the proposed method reduces computational time by one to two orders of magnitude compared to existing methods. 
%Numerical studies on an IEEE test system demonstrates the effectiveness of the proposed method compared to the state of the art framework.\\
\end{abstract}


\section{Introduction}
%A higher need for situational awareness 

%As %a direct consequence of 
%The growing integration of DERs (distributed energy sources) and demand-side technologies introduces a high level of uncertainties to distribution systems, which in turn also increases the need for situational awareness of distribution system. % and DERs proliferation of distributed energy resources, electric vehicles  and controllable loads  on the distribution grid, has led to an increase in the installation of monitoring equipment. This growing availability of high-precision and high sample-rate measurements on the distribution grid has provided justification for research on data-driven topology and network admittance information identification. 

An accurate distribution network model is crucial for implementing demand-side technologies and monitoring distributed energy resources (DERs). However, these models are often unavailable
or outdated due to the continuous integration of DERs and frequent reconfigurations. On the other hand, the growing availability of high-precision %and high sample-rate 
measurements on the distribution grid provides a unique opportunity to develop data-driven real-time identification methods. 

Many data-driven techniques %have been 
were proposed in the literature to identify topology \cite{wang2017pmu,6760120,9151363}  or estimate line parameters \cite{6672179,9960997} or both \cite{8122027,8581489,9729419,8798652}. The authors in \cite{wang2017pmu} performed topology change detection by estimating the dynamic Jacobian and system state matrices. The sparse Markovian random field property of grid voltage magnitude measurements were exploited in  \cite{6760120} to reconstruct the topology of a portion of the distribution grid using measurements from phasor measurement units (PMUs). An offline total least squares method was proposed in \cite{6672179} to estimate the positive sequence line parameters of transmission lines while a physics-informed graphical learning method was developed in \cite{9960997}. However, the algorithm in \cite{9960997} requires initial knowledge of the topology and line parameters. A combined framework for parameter and topology estimation was introduced in \cite{8122027}, with enhancements to account for state changes in the distribution network developed in \cite{8581489}. %A joint parameter and topology estimation framework was proposed in \cite{8122027} and an improvement on this method considering state changes in the distribution network was developed in \cite{8581489}. 
An alternating direction method of multipliers (ADMM)-based framework where both smart meter and PMU measurements were used in the joint estimation of line parameters and topology identification was proposed in \cite{9729419}.

While various algorithms have been proposed, many encounter specific limitations, such as difficulties in handling Gaussian noise in measurements \cite{9151363,8601410}, issues with high dimensionality, and a significant computational burden for large networks \cite{8601410,8798652}. Additionally, many algorithms are not directly applicable to unbalanced distribution grids \cite{9027950,8122027,8581489}. As highlighted in a recent tutorial \cite{10113752}, only about 10\% of papers on distribution grid identification focus on multi-phase distribution networks. Furthermore, most approaches have not addressed the added uncertainties introduced by intermittent DERs such as wind and solar PV.

%While various algorithms have been proposed, some of them suffer from shortcomings, particularly in handling Gaussian noise present in the measurements \cite{9151363,8601410}, high dimensionality issues and  high computational burden for larger networks \cite{8601410,8798652} and inability to directly implement them on unbalanced distribution grids \cite {9027950,8122027,8581489}. %For example in a recent published tutorial on "learning" distribution grids \cite{10113752}, 
%Indeed, as mentioned in the recent tutorial \cite{10113752}, only about $10 \%$ of the papers on distribution grid identification %discussed 
%focused on multiphase distribution networks. %\cite{10113752}.  
%Also, most have failed to consider estimation in the presence of added uncertainties due to the presence of intermittent distributed energy resources such as wind and solar PV. 

In this paper, we propose a two-stage line estimation method for multiphase unbalanced distribution networks leveraging the regression theorem of multivariate Ornstein-Uhlenbeck and Broyden
diagonal elements analysis. %theorem-based regression method for the estimation of line parameters in multiphase unbalanced distribution systems. 
%\color{red}
Comprehensive simulation studies using real-life load and PV data show that the proposed method outperforms the Lasso 
\cite{8273895} 
and adaptive Lasso 
\cite{8601410} in estimating line susceptance and is comparable to Lasso and adaptive Lasso in estimating line conductances. 
%lower noise levels and is comparable with the adaptive lasso
%method at higher noise levels. Nevertheless, 
Importantly, the proposed method requires two orders of magnitude
less computational time than adaptive Lasso, making it more applicable in online applications. \color{black} %can efficiently estimate the parameters in the presence of measurement noise and the stochastic nature of renewable power generation. %Also, leveraging on the correctly estimated admittance matrix, an online algorithm is implemented to detect critical events that are capable of altering the admittance matrix.
%Comprehensive simulation studies using real load and PV data show that the proposed method outperforms the Lasso method \color{blue}\cite{8273895} \color{black} at lower  noise levels and is comparable with the adaptive Lasso method \cite{8601410} at higher noise levels. Nevertheless, the proposed method requires \color{blue} approximately ${\frac{1}{100}}^{th}$ of the computation time for adaptive Lasso making it more applicable in online applications \color{black}.

\section{The Modeling of Multi-phase and Unbalanced Distribution Systems}
%\color{red} give the compact power flow model (show delta, V, P Q are coupled in a nolinear way) and needs the Y matrix. Then focus on explaining the Y matrix (complexity compared to the transmission system counterpart)\color{black} 
%A method is proposed to estimate the line parameters of multiphase unbalanced distribution networks using real-time measurements of nodal power injections $(P_{i},Q_{i})$, and $\mu$-pmu measurements of complex nodal voltages $(V_{i})$. 
A multiphase unbalanced distribution grid can be modeled as an undirected graph  $\mathscr{G}=(\mathscr{R}, \mathscr{S})$, where $\mathscr{R}=\{0,1,2, \ldots, N\}$ denotes the set of buses/nodes and the set $\mathscr{S}=\{(i, j), i, j \in \mathscr{R}\}$ represents the branches. The number of branches between any two nodes can be $1 \leq \card {\mathscr{S}}\leq 3$ (depending on the phase number).  \par
%\color{blue}
Let $Z_{i j}$ be the three-phase impedance matrix representing line impedance of the branch $(i, j)$  and $z_{i j}^{n p}$ be the impedance entries in the matrix $Z_{i j}\left(\forall n \in \alpha_{i} \subseteq\{a, b, c\}\right)$ and $\left(\forall p \in \alpha_{j} \subseteq\{a, b, c\}\right)$.  %\color{red} %you use n to denote phase in later sections. can we make it consistent here? 
%Can I replace alpha-i by n
%alpha-j by p in (1)?
%and (2)?
%I don't see them being used elsewhere
%\color{black}

%  \small
% \begin{equation}
%       \begin{aligned}
%         v_{j}^{\alpha_{j}} & =\left[\begin{array}{lll}
%         V_{j}^{a}& V_{j}^{b} &V_{j}^{c}
%         \end{array}\right]^{T}\left[\begin{array}{lll}
%         \left(V_{j}^{a}\right)^{*} & \left(V_{j}^{b}\right)^{*} & \left(V_{j}^{c}\right)^{*}
%         \end{array}\right] \\
%         % \forall j \in \mathcal{N}  \\
%     \end{aligned} \label{eqn:PF2}
%  \end{equation}
%   \normalsize
% \small
% \begin{equation}
%       \begin{aligned}
%         S_{i j} & =\left[\begin{array}{lll}
%         V_{j}^{a}& V_{j}^{b} &V_{j}^{c}
%         \end{array}\right]^{T}\left[\begin{array}{lll}
%         \left(I_{i j}^{a}\right)^{*} & \left(I_{i j}^{b}\right)^{*} & \left(I_{i j}^{c}\right)^{*}
%         \end{array}\right]\\
%     \end{aligned} \label{eqn:PF3}
%  \end{equation}
%  \normalsize
$\alpha_{j}$ denotes the set of phases of bus $j,  \forall j \in \mathcal{N}$ and $\alpha_{i j}$ denotes the set of phases in branch $(i, j)$, $\forall(i, j) \in \mathscr{S}$. Similarly, let  $S_{i j}$ and $I_{i j}^{\alpha_{i j}}$ be the complex power flow matrix and complex current flow in branch $(i, j) \in \mathscr{S}$ from bus $i$ to bus $j$. Then, the complex voltage and  power flow matrices are expressed as \cite{7317598}: %\color{red} what does H mean in (1)? \color{black}
\small
 \begin{equation}
      \begin{aligned}
        v_{j}^{\alpha_{j}} & =v_{i}^{\alpha_{i}}-\left(Z_{i j} S_{i j}^{H}+S_{i j} Z_{i j}^{H}\right) \\
    \end{aligned} \label{eqn:PF1}
 \end{equation}
 \normalsize
 where $v_{j}^{\alpha_{j}}=[V_{j}^{a} V_{j}^{b} V_{j}^{c}]^{T}[(V_{j}^{a})^{*}  (V_{j}^{b})^{*}  (V_{j}^{c})^{*}]$, $S_{i j}  =[V_{j}^{a} V_{j}^{b} V_{j}^{c}
        ]^{T}[
        (I_{i j}^{a})^{*}  (I_{i j}^{b})^{*}  (I_{i j}^{c})^{*}
        ]$ and $*^H$ is the hermitian notation. \color{black} Let $\lambda  :=$ $\angle V_{j}^{a}/\angle V_{j}^{b} =\angle V_{j}^{b}/\angle V_{j}^{c}=\angle  V_{j}^{c}/\angle V_{j}^{a}=e^{j 2 \pi / 3}$, $\boldsymbol{\lambda}=\left[\begin{array}{ll}1 & \lambda\end{array} \lambda^{2}\right]^{\top}$, $\lambda^{2}=\lambda^{*}=-\frac{1}{2}+j \frac{\sqrt{3}}{2}$ and:
\small
\begin{equation}
    \begin{aligned}
{{Z}}_{ij}=\operatorname{diag}\left(\boldsymbol{\lambda}^{*}\right) \tilde{Z}_{ij} \operatorname{diag}(\boldsymbol{\lambda})=\left[\begin{array}{ccc}
z_{ij}^{a,a} & \lambda^{*} z_{ij}^{a,b} & \lambda z_{ij}^{a,c} \\
\lambda z_{ij}^{b,a} & z_{ij}^{b,b} & \lambda^{*} z_{ij}^{b,c} \\
\lambda^{*} z_{ij}^{c,a} & \lambda z_{ij}^{c,b} & z_{ij}^{c,c}
\end{array}\right]
    \end{aligned}
\end{equation}
\normalsize
Following the assumptions made in \cite{7317598}, %Hence, 
equation %\color{red} some brackets are missing, check other places as well \color{black} 
(\ref{eqn:PF1}) can be further expressed as\color{black}:
% \small
% \begin{equation}
%         \begin{aligned}
% & v_{j}=v_{i}-r_{n}^{ij} \bar{P}_{i j}-r_{n}^{ij} \bar{Q}_{i j} \quad \forall j \in N  \\
%    \end{aligned} \label{eqn:PF5}
% \end{equation}\par
% \normalsize
\small
\begin{equation}
\mathbf{v}_{j}^{\alpha_{j}}-\mathbf{v}_{i}^{\alpha_{i}}=2 \operatorname{Re}\left[\operatorname{diag}\left(\boldsymbol{\lambda} \boldsymbol{\lambda}^{H} \operatorname{diag}\left({S}_{ij}\right) \tilde{Z}_{ij}^{H}\right)\right]
\end{equation}
\normalsize
which describes that voltage drop in each phase depends on the power flows in its own phase and those of adjacent phases due to mutual coupling amongst phases. 
% where  $\tilde{Z}_{n}=\operatorname{diag}\left(\boldsymbol{\lambda}^{*}\right) {Z}_{n} \operatorname{diag}(\boldsymbol{\lambda})$. 
%\normalsize \color{red} what 's the relation between $\tilde{Z}_{ij}$ and $Z_n$, $z_n^{ij}$?\color{black}

Our main focus is the $3 \times 3$ complex impedance matrices ${{Z}}_{ij}$ %\color{red} why is it in bold?Also, why is there a tilde? \color{black} 
and with $z_{ij}^{n,p}=r_{ij}^{n,p}+j x_{ij}^{n,p}$ being the $(n, p)$ entry, $\forall n,p \subseteq\{a, b, c\} $ of ${Z}_{ij}$. %\color{red} what is tilde Z? \color{black}
% From the symmetry of $\tilde{Z}_{n}$ and if we let $\lambda^{2}=\lambda^{*}=-\frac{1}{2}+j \frac{\sqrt{3}}{2}$, matrix ${{Z}}_{n}$ can be defined as:

Then the admittance matrix ${{Y}}_{ij}={{Z}}_{ij}^{-1}$ and for every branch ${ij}$ in the network: %\color{red} what's the relation between $Y_n$ and $Y_{ij}^{n,p}$. Are they the same thing? If so, why not only keep (7)?\color{black}
% \begin{equation}
%     \begin{aligned}
% \mathbf{Y_{n}} = {\mathbf{Z}}^{-1}_{n}=\left[\begin{array}{ccc}
% y_{n}^{a,a} & y_{n}^{a,b} &  y_{n}^{a,c} \\
% y_{n}^{b,a} & y_{n}^{b,b} &  y_{n}^{b,c} \\
% y_{n}^{c,a} &  y_{n}^{c,b} & y_{n}^{c,c}
% \end{array}\right]
%     \end{aligned}\label{eqn:admittance}
% \end{equation}
\small
\begin{equation}
    \begin{aligned}
{Y_{ij}} =  {G}_{ij}+j{B}_{ij}=\left[\begin{array}{ccc}
G_{ij}^{a,a} & G_{ij}^{a,b} &  G_{ij}^{a,c} \\
G_{ij}^{b,a} & G_{ij}^{b,b} &  G_{ij}^{b,c} \\
G_{ij}^{c,a} & G_{ij}^{c,b} &  G_{ij}^{c,c} 
\end{array}\right] +\\
j\left[\begin{array}{ccc}
B_{ij}^{a,a} & B_{ij}^{a,b} &  B_{ij}^{a,c} \\
B_{ij}^{b,a} & B_{ij}^{b,b} &  B_{ij}^{b,c} \\
B_{ij}^{c,a} & B_{ij}^{c,b} &  B_{ij}^{c,c} 
\end{array}\right]
    \end{aligned}\label{eqn:conductance}
\end{equation}
\normalsize
Where the admittance matrix $Y_{ij}^{n,p}$ is a full matrix when all three lines connecting buses $i$ and $j$ are present  and there exists coupling amongst the phases. Conversely, when  some lines are missing, the enteries for the admittance values in equation (\ref{eqn:conductance}) \color{black}for the missing lines   are $\mathbf{0}$ . This generally leads to a less sparse admittance matrix with more prominent off-diagonal values as compared to transmission systems. The added complexities of missing lines and coupling amongst phases contribute to why most of the line parameter estimation  algorithms in the literature which are geared towards transmission systems and single-phase models of distribution systems cannot be trivially extended to the multiphase distribution system. In the next section (\ref{Dynmodel}) , we explain how a representative load model can be leveraged to estimate the elements of the admittance matrix in multiphase distribution systems.

\section{Problem Formulation}

\subsection{Dynamic Model of the Load}\label{Dynmodel}
%\color{red} Move the dynamic load model to the place after discussing Y matrix. give both (9)-(10) with power flow, note the phase (diff for distribution system),  Justify the dynamic load model, and also mention you will use real load data in simulation \color{black} \par

% In this work we use a dynamic load model introduced in \cite{7095616}  and successfully  used in  \cite{9281825,du_pierrou_wang_kassouf_2021} to describe a wide range of loads ranging from thermostatically controlled loads, power electronic converters, induction motors, combined impacts of otherwise unmodeled distribution load tap changer (LTC) transformers, etc
In this work we use a dynamic load model introduced in \cite{7095616}  and successfully  used in  \cite{du_pierrou_wang_kassouf_2021} to describe a wide range of loads ranging from thermostatically controlled loads, power electronic converters, induction motors, combined impacts of otherwise unmodeled distribution load tap changer (LTC) transformers, etc.  The range of time constants is wide, ranging from cycles to minutes, and is therefore justifiably adequate to represent the real load data implemented in the simulations for this paper. This load model which is assumed to be perturbed by independent Gaussian variations is defined in detailed form \cite{7095616}:
\small
\begin{equation}
  \dot{\delta_{i}^{n}} =\frac{1}{\tau_{ppi}^{n}}(P_{i}^{s,n}(1+\sigma_{i}^{pp,n}\xi_{i}^{pp,n})-P_{i}^{n})\label{eq:angleqnpert}
, \forall n \in\{a, b, c\}\end{equation}
\normalsize
\small
\begin{equation}
  \dot{V_{i}^{n}} =\frac{1}{\tau_{qqi}^{n}}(Q_{i}^{s,n}(1+\sigma_{i}^{qq,n}\xi_{i}^{qq,n})-Q_{i}^{n})\label{eq:volteqnpert}
, \forall n \in\{a, b, c\} \end{equation}
\normalsize

\small
\begin{equation}
\begin{aligned}
P_{i}^{n}= & V_{i}^{n} \sum_{p \in \Omega} \sum_{n=1}^{N} V_{j}^{p}\left[G_{i j}^{n, p} \cos \left(\delta_{j}^{n}-\delta_{j}^{p}\right)+\right. \\
& \left.B_{ij}^{n, p} \sin \left(\delta_{j}^{n}-\delta_{j}^{p}\right)\right]  \\
Q_{i}^{n}= & V_{i}^{n} \sum_{p \in \Omega} \sum_{n=1}^{N} V_{j}^{p}\left[G_{i j}^{n, p} \sin \left(\delta_{i}^{n}-\delta_{j}^{p}\right)-\right. \\
& \left.B_{ij}^{n, p} \cos \left(\delta_{i}^{n}-\delta_{j}^{p}\right)\right]
\end{aligned}
\end{equation}
\normalsize
%\color{red} Is this model for a particular 3-phase node? what if different nodes? for example, the a phase of node A and the b phase of node B? if node A and B are connected? Also, I don't see how this power flow and the power flow discussed in the previous section are related? \color{black}
where $\delta_{i}$ and  $V_{i}$ are the voltage angle and the voltage magnitude respectively of bus, $i$ , $P_{i}^{n}$ and $Q_{i}^{n}$ are the active and reactive power injection of node $i$ respectively at phase $n$, $\tau_{ppi}^{n}$ and $\tau_{qqi}^{n}$ are the active and reactive power time constants respectively, $\xi_{i}^{pp,n}$ and $\xi_{i}^{qq,n}$ are standard Gaussian random variables, $\sigma_{i}^{pp,n}$ and $\sigma_{i}^{qq,n}$ are the noise intensities of the load variations at node $i$ and phase $n$.
Static loads can be represented by applying the limit $\tau_{ppi}^{n}, \tau_{qqi}^{n} \to 0$  and distributed generation such as PVs are modeled as negative dynamic loads.

%\color{red}In the simulation studies of this paper, real load data and PV data will be used ...\color{black} 
In the simulation studies of this paper, real load and PV data will be used to justify the defined load model's replication of real-time load fluctuations and changing grid conditions. In section (\ref{ouprocess}), a multivariate Ornstein-Uhlenbeck regression theorem is proposed to compactly describe the line parameter identification problem. 
 
\subsection{Multivariate OU Regression Theorem}\label{ouprocess}
The compact form of the dynamic load model (\ref{eq:angleqnpert}-\ref{eq:volteqnpert})can be rewritten as a  multivariate Ornstein-Uhlenbeck process:
\small
\begin{equation}
\dot{\mathbf{x}} = A \mathbf{x} + B \mathbf{\xi} \label{eq:compacteqn}
\end{equation}
\normalsize
%\color{blue}
\small
\begin{equation}
\begin{aligned}
    \underbrace{\begin{bmatrix} \dot{\bm{\delta}}_{i} \\ \dot{\mathbf{V}}_{i}\end{bmatrix}}_{\dot{\mathbf {x}}} = 
\underbrace{\begin{bmatrix} T_{p}^{-1} &  \\  & T_{q}^{-1} \end{bmatrix} 
\begin{bmatrix} J_{P\delta} & J_{PV} \\ J_{Q\delta} & J_{QV} \end{bmatrix}}_{\textbf{A}}
\underbrace{\begin{bmatrix} \bm{\delta}_{i} \\ \mathbf{V}_{i}  \end{bmatrix}}_{\textbf{x}} + \\
&\quad \\
\underbrace{\begin{bmatrix} T_{p}^{-1} P^{s}\sum^{p} & 0 \\ 0 & T_{q}^{-1} P^{s}\sum^{q} \end{bmatrix} }_{\textbf{B}}
\underbrace{\begin{bmatrix} \bm{\xi}^{pp}_{i} \\ \bm{\xi}^{qq}_{i}\end{bmatrix}}_{\mathbf{\xi}}
\end{aligned} \label{eq:PQeqnpert}
\end{equation}
\normalsize
\color{black}
%\color{red} no n for delta and V on the left hand side of (12)? also, delta should be in bold in (12) and in the explanations.\color{black} 
\text{where} $\mathbf{x} = \begin{pmatrix} \bm{\delta }\\ \bm{V} \end{pmatrix}$,$J_{\mathbf{P}, \mathbf{V}} = \dfrac{\partial \mathbf{P}}{\partial \mathbf{V}}$, $J_{\mathbf{P}, \mathbf{\bm{\delta}}} = \dfrac{\partial \mathbf{P}}{\partial \bm{\delta}}$, $J_{\mathbf{Q}, \mathbf{V}} = \dfrac{\partial \mathbf{Q}}{\partial \mathbf{V}}$, $J_{\mathbf{Q}, \bm{\delta}} = \dfrac{\partial \mathbf{Q}}{\partial \bm{\delta}}$ ,%\color{blue}
${A}$ \color{black}is the system state matrix and $T_{p},T_{q}$ are diagonal matrices representing the active and reactive load power constants, respectively \color{black}. The system state matrix $A$ corresponds to a scaled Jacobian matrix that contains important information about the dynamic load time constants and line parameters, ${G}_{ij}$ and ${B}_{ij}$. The first step is to estimate the state matrix $A$, and extract the Jacobian matrices $J_{\mathbf{P}, \mathbf{V}},J_{\mathbf{P}, \mathbf{\bm{\delta}}}, J_{\mathbf{Q}, \mathbf{V}}$, and $J_{\mathbf{Q}, \bm{\delta}}$ from $A$. The next step will be to use the  nonlinear partial differential equations relating power injections to the voltage magnitude and angle to estimate the ${Y}_{ij}= {G}_{ij}+ {B}_{ij}$. %\color{red} do you need to put Y, G, B in bold? I think some places, they are bold and some are not.\color{black} 
In the next section (\ref{ouprocess2}), these extraction and estimation processes are discussed.  \par
%\color{red} relate the matrix to Y matrix.  \color{black}
%\color{red} add connecting sentences that our goal is to first estimate A, and then the Jacobian matrix, and then the Y matrix \color{black}

%\color{red}A new section for methodology\color{black}

\section{The Proposed Two-Stage Line Estimation Method} %Based on the Ornstein-Uhlenbeck Process} %Methodology}
\subsection{Estimating the Load and Line Parameters from PMU Measurements }\label{ouprocess2}
%\color{red} the OU process property 
Leveraging the regression theorem of the multivariate OU process \cite{gardiner_2009_handbook}, %\color{red} cite the handbook \color{black},
which posits that if there exist sufficient measurements that the  $\tau$-lag correlation matrix $C(\Delta t)$ can be estimated  %calculated 
and the power system is operating in a normal steady state, the  system's stable state matrix $A$ can be numerically estimated as follows \cite{du_pierrou_wang_kassouf_2021}:
\small
\begin{equation} \label{eq:Amatrix}
\hat{A}=\frac{1}{\Delta t} \ln \left[\hat{C}(\Delta t) \hat{C}(0)^{-1}\right] 
\end{equation}
\normalsize
The sample covariance matrices $\hat{C}(0),\hat{C}(\Delta t)$ are calculated from a finite data set as follows:
\small{
\begin{gather}\label{covmatrices}
\hat{C}(0)=\frac{1}{S-1}\left(F_{1: S}-\hat{\boldsymbol{\mu}}_{x} \mathbf{1}_{1: S}\right)\left(F_{1: S}-\hat{\boldsymbol{\mu}}_{x} \mathbf{1}_{1: S}\right)^{T} \\
\hat{C}(\Delta t)=\frac{1}{S-1}\left(F_{1+K: S}-\hat{\boldsymbol{\mu}}_{x} \mathbf{1}_{1: S-K}\right)\left(F_{1: S-K}-\hat{\boldsymbol{\mu}}_{x} \mathbf{1}_{1: S-K}\right)^{T} 
\end{gather}
\normalsize}
and the sample mean $\hat{\mu}$ is calculated as:
\small
\begin{equation}\label{samplemean}
    \hat{\boldsymbol{\mu}}_{x}=\frac{1}{S} \sum_{i=1}^{S} \boldsymbol{x}^{(i)} 
\end{equation}
\normalsize
%\color{red} no space before ``where''. check throughout the paper\color{black}
where $S$ is the sample size, $\Delta t$ is the sampling time step, $F=\left[\boldsymbol{x}_{1}, \boldsymbol{x}_{2}, \cdots, \boldsymbol{x}_{S}\right]$ is a $R_{s} \times$ $S$ matrix assuming there are $R_{s}$ state variables, $K$ is the number of samples that correspond to the selected time lag and  $F_{i: j}$ represents the submatrix of $F$ from $i$ to $j$ columns, $\mathbf{1}_{S}$ is an $S$ by $1$ vector of ones. Note that $\bm{x}$ are voltage magnitudes and phase angles that can be collected from micro-PMUs. \par
%For an online estimation implementation, the recursive method in \cite{8718822} is used  to recursively update the parameters of the covariance matrices so as not to repeatedly recalculate the basic equations. 
% Once the $\mathbf{A}$ matrix has been successfully estimated from (\ref{eq:Amatrix}), the Jacobian matrices can be extracted using the weighted least square (WLS) regression method \cite{monticelli1999state} to calculate  the load time constants and finally, the initial values $\mathbf{G}_{ij}^{*}$ and $\mathbf{B}_{ij}^{*}$.
Once the $\mathbf{A}$ matrix has been successfully estimated from (\ref{eq:Amatrix}), the Jacobian matrices can be extracted using the weighted least square (WLS) regression method to calculate  the load time constants and finally, the initial values $\mathbf{G}_{ij}^{*}$ and $\mathbf{B}_{ij}^{*}$. Equation \eqref{eq:tauest} is derived from equation \eqref{eq:angleqnpert} and $\frac{1}{\tau^{n}_{pp_{i}}}$ is then calculated using WLS. Similarly, $\frac{1}{\tau^{n}_{qq_{i}}}$ is also calculated from derivations of equation \eqref{eq:volteqnpert}.
%\color{red} taupp and tau qq should also have n(phase) \color{black}
\small
\begin{equation} \label{eq:tauest}
\underbrace{\left[\begin{array}{l}
\frac{1}{\Delta t}\left(\delta_{i}^{n,(2)}-\delta_{i}^{n,(1)}\right)  \\
\cdots \\
\frac{1}{\Delta t}\left(\delta_{i}^{n,(k)}-\delta_{i}^{n,(k-1)}\right)
\end{array}\right]}_{U}=\underbrace{\left[\begin{array}{l}
\hat{\mu}_{P_{i}}^{n}-P_{i}^{n,(1)} \\
\cdots \\
\hat{\mu}_{P_{i}}^{n}-P_{i}^{n,(k-1)}
\end{array}\right]}_{L} \underbrace{\left[\frac{1}{\tau^{n}_{pp_{i}}}\right]}_{\beta}
\end{equation}
 Where $P_{i}^{(n,(k))}$ represents the $k^{\text {th }}$ observation of active power of the bus $i$ and phase $n$ ;$\Delta t$ is the time lag, $\hat{\mu}_{P_{i}}^{n}$ denotes the sample mean of $P_{i}^{n}, \delta_{i}^{(n,(k)}$ represents the $k^{\text {th }}$ observation of voltage angle of bus $i$ and phase $n$. $\forall n \in\{a, b, c\}$, using WLS regression, equation (\ref{eq:tauest}) is solved by $\boldsymbol{\beta}=\left(L^{T} W L\right)^{-1} L^{T} W \boldsymbol{U}$ with $W$ representing the weight matrix. In this paper, we set $W= \mathbb{I}^{n \times n}$. %\color{red} what weight matrix did you choose? identity matrix? \color{black} 
 Once the time constants $\hat{\tau}_{pp_{i}}$ and  $\hat{\tau}_{qq_{i}}$ have been estimated, the estimated derivatives  $\hat{J}_{P_{i}\delta_{j}}$, $\hat{J}_{P_{i}V_{j}}$,$\hat{J}_{Q_{i}\delta_{j}}$, $\hat{J}_{Q_{i}V_{j}}$, $\forall{(i,j)}$ can be obtained from $\hat{A}$. Taking the derivatives of \eqref{eq:PQeqnpert} will result in:

% \begin{equation} \label{eq:GBest}
%     \begin{bmatrix} \hat{J}_{P_{i}\delta_{j}} \\  \hat{J}_{P_{i}V_{j}} \\ \hat{J}_{Q_{i}\delta_{j}}\\ \hat{J}_{Q_{i}V_{j}} \end{bmatrix} = \begin{bmatrix} V_{i}V_{j} \sin{\delta_{ij}}& -V_{i}V_{j} \cos{\delta_{ij}}\\  V_{i} \cos{\delta_{ij}} & V_{i} \sin{\delta_{ij}} \\ -V_{i}V_{j} \cos{\delta_{ij}} &  V_{i}V_{j} \sin{\delta_{ij}} \\ V_{i} \sin{\delta_{ij}} &  V_{i} \cos{\delta_{ij}} \end{bmatrix} \begin{bmatrix} G_{ij}^{*} \\ B_{ij}^{*} \end{bmatrix}
% \end{equation}\par
\small{
\begin{equation}\label{eq:GBest}
\underbrace{\begin{bmatrix} \hat{J}_{P^{n}_{i}\delta_{j}^{p}} \\  \hat{J}_{P^{n}_{i}V_{j}^{p}} \\ \hat{J}_{Q^{n}_{i}\delta_{j}^{p}}\\ \hat{J}_{Q^{n}_{i}V_{j}^{p}} \end{bmatrix}}_{U} = \\ 
\underbrace{\begin{bmatrix} V_{i}^{n}V_{j}^{p} \sin{(\delta_{ij}^{n,p})}& -V_{i}^{n}V_{j}^{p} \cos{(\delta_{ij}^{n,p})}\\  V_{i}^{n} \cos{(\delta_{ij}^{n,p})} & V_{i} \sin{(\delta_{ij}^{n,p})} \\ -V_{i}^{n}V_{j}^{p} \cos{(\delta_{ij}^{n,p})} &  V_{i}^{n}V_{j}^{p} \sin{(\delta_{ij}^{n,p})} \\ V_{i}^{n} \sin{(\delta_{ij}^{n,p})} &  V_{i}^{n} \cos{(\delta_{ij}^{n,p})} \end{bmatrix}}_{L} \underbrace{\begin{bmatrix} G_{ij}^{*^{n,p}} \\ B_{ij}^{*^{n,p}} \end{bmatrix}}_{\beta}
\end{equation}
}
\normalsize
where $\delta_{ij}^{n,p} =\delta_{i}^{n}-\delta_{j}^{p}$. WLS  regression method $(\boldsymbol{\beta}=\left(L^{T} W L\right)^{-1} L^{T} W \boldsymbol{U})$ is again applied to equation (\ref{eq:GBest}) to estimate the initial line parameters value $\mathbf{G}_{ij}^{{*},{n,p}}$ and $\mathbf{B}_{ij}^{{*},{n,p}}$.

%\color{red} mention this has been done in transmission network in Mingqiu's paper. But it works only for the diagonal elements. However, the accuracy for distribution system is not good enough even for diagonal element for due to the complexity and scale. So, you propose a two-stage method, treat.. as initial values, to further estimate the whole Y matrix including both diagonal and off diagonal elements \color{black}
The Ornstein-Uhlenbeck regression theorem has previously been%proposed 
applied to the estimation of line parameters for transmission systems in \cite{du_pierrou_wang_kassouf_2021}. However, the method cannot estimate all elements of the admittance matrix well for a multiphase unbalanced distribution network due to %the complexities. %this paper demonstrates that the method cannot be trivially extended to multiphase distribution systems due to properties such as 
unbalanced loading and coupling amongst phases.  %which has been discussed in previous sections. 
In this paper, we will extend the method by adding a second stage %final identification method 
to improve the accuracy of the parameter identification method. This extension is discussed in section (\ref{Broyden})

\subsection{The Second Stage of Parameter %Evaluation 
Estimation Using Broyden Diagonal Elements Analysis}\label{Broyden}
%\color{red} need to extend this part \color{black}
In this section, the approximate values of conductances and susceptances obtained in (\ref{ouprocess2}) are further used to obtain estimates with a higher accuracy. Using the available active and reactive power injection measurements, at the buses with measurement devices, the change in the active and reactive power matrices are built as \cite{9027950}: %\color{red} Please check your P, Q, G, B matrices throughout the paper. Should they be in bold or not? \color{black} \color{blue}
\small{
\begin{equation} \label{eq:PQdelta}
\left[\begin{array}{c}
\Delta \mathbf{P} \\
\Delta \mathbf{Q}
\end{array}\right]= \left[\begin{array}{lll}
\frac{\partial {P}}{\partial G} & \frac{\partial {P}}{\partial {B}} & \frac{\partial {P}}{\partial {\delta}} \\
\frac{\partial {Q}}{\partial {G}} & \frac{\partial {Q}}{\partial {B}} & \frac{\partial {Q}}{\partial {\delta}}
\end{array}\right]  
\times\left[\begin{array}{c}
\Delta \mathbf{G} \\
\Delta \mathbf{B} \\
\Delta {\bm{\delta}}
\end{array}\right]
\end{equation}
\begin{equation} \label{eq:fine_est}
\left[\begin{array}{c}
\Delta \mathbf{G} \\
\Delta \mathbf{B} \\
\Delta {\bm{\delta}}
\end{array}\right]=\underbrace{\left[\begin{array}{lll}
\frac{\partial {P}}{\partial {G}} & \frac{\partial{P}}{\partial B} & \frac{\partial {P}}{\partial {\delta}} \\
\frac{\partial {Q}}{\partial {G}} & \frac{\partial f{Q}}{\partial {B}} & \frac{\partial {Q}}{\partial {\delta}}
\end{array}\right]^{\dagger}}_{\textbf{J}^{-1}} \cdot\left[\begin{array}{c}
\Delta \mathbf{P} \\
\Delta \mathbf{Q}
\end{array}\right]
\end{equation}
\color{black}

\begin{equation} \label{eq:updateG}
\left[\begin{array}{l}
\bm{G} \\
\bm{B}
\end{array}\right]^{(k+1)}=\left[\begin{array}{l}
\bm{G} \\
\bm{B} 
\end{array}\right]^{(k)}+\left[\begin{array}{c}
\Delta \bm{G} \\
\Delta \bm{B} 
\end{array}\right]
\end{equation}

\begin{equation}
    \begin{aligned}
 \frac{\partial \bm{P}_{i}^{n}}{\partial {G}_{ij}^{n,p}} = \frac{\partial \bm{Q}_{i}^{n}}{\partial {B}_{ij}^{n,p}}=u_{ij}\left(({V}_{i}^{n})^{2}-{V}_{i}^{n} {V}_{j}^{p} \cos \left({\delta}_{i}^{n}-{\delta}_{j}^{p}\right)\right) \\
 \frac{\partial \bm{P}_{i}^{n}}{\partial {B}_{ij}^{n,p}}=\frac{\partial \bm{Q}_{i}^{n}}{\partial {G}_{ij}^{n,p}}=-u_{ij}\left(\bm{V}_{i}^{n} \bm{V}_{j}^{p} \sin \left({\delta}_{i}^{n}-{\delta}_{j}^{p}\right)\right) 
    \end{aligned}
\end{equation}
\normalsize}
%\color{red} I think if given in equation like (22)-(24), phases should be denoted. (Unlike (12), I can accept that phases are not denoted as they are in compact matrix form) \color{black}
where %\color{blue} 
in (\ref{eq:updateG}), 
$\bm{G}=\mathrm{vec}(G_{ij}^{n,p})$,$\bm{B}=\mathrm{vec}(B_{ij}^{n,p})$, , $u_{ij}$ is the connectivity indicator for branch ${ij}$, $u_{ij}=1$ when node ${i}$ is connected to node ${j}$ and $u_{ij}=0$ otherwise. 
%\color{red} Why are $V_i^n$, other voltages, $\delta_i^p$ in bold? \color{black}

%\color{blue}
\small{
\begin{equation}
    % \begin{align}
\frac{\partial {P}_{i}^{n}}{\partial {\delta}_{i j}} \\
=\left\{\begin{array}{l}
{V}_{i}^{n} \sum_{j, j \neq i} {V}_{j}^{p}\left(-G_{i j}^{n,p}\sin \left({\delta}_{i}^{n}-{\delta}_{j}^{p}\right)+
B_{i j}^{n,p} \cos \left({\delta}_{i}^{n}-{\delta}_{j}^{p}\right)\right),\\
i=j, \forall  n,p \in \{a,b,c\}\\
{V}_{i}^{n} {V}_{j}^{p}\left(G_{i j} ^{n,p}\sin \left({\delta}_{i}^{n}-{\delta}_{j}^{p}\right)-B_{i j}^{n,p} \cos \left({\delta}_{i}^{n}-{\delta}_{j}^{p}\right)\right), \\
i \neq j, \forall  n,p \in \{a,b,c\}
\end{array}\right. 
% \end{align}
\end{equation}

\begin{equation}
% \begin{align}
\frac{\partial {Q}_{i}^{n}}{\partial {\delta}_{i j}} \\
 =\left\{\begin{array}{l}
{V}_{i}^{n} \sum_{j, j \neq i} {V}_{j}^{p}\left(G_{i j}^{n,p} \cos \left(\delta_{i}^{n}-\delta_{j}^{p}\right)+
B_{i j}^{n,p} \sin \left({\delta}_{i}^{n}-{\delta}_{j}^{p}\right) \right), \\
i=j, \forall  n,p \in \{a,b,c\} \\
-{V}_{i}^{n} {V}_{j}^{p}\left(G_{i j} ^{n,p}\cos \left({\delta}_{i}^{n}-{\delta}_{j}^{p}\right)+B_{i j}^{n,p} \sin \left({\delta}_{i}^{n}-{\delta}_{j}^{p}\right)\right),\\
i \neq j, , \forall  n,p \in \{a,b,c\}
\end{array}\right.
% \end{align}
\end{equation}
\color{black}
\normalsize}
% Next, a quasi-Newton-Raphson method also known as the Broyden method \cite{Remani_2012} is  used 
Next, a quasi-Newton-Raphson method also known as the Broyden method is  used 
to solve equation \eqref{eq:fine_est} for $\Delta G$, $\Delta B$ and the line parameters are updated using equation \eqref{eq:updateG}. In contrast to the Newton-Raphson method used in\cite{9027950}, Broyden's method was introduced to solve the system of equations as it is designed to  improve Newton's method with respect to storage and approximation of the Jacobian. This is advantageous for estimating line parameters in distribution systems with a large number of nodes. However, the price paid for such savings is the reduction in convergence from quadratic to superlinear.
In summary, a representative load model is reformulated as a multivariate OU model with measurement noise modeled as Gaussian ($\sigma$ and $\xi$ in equations (\eqref{eq:angleqnpert})-(\eqref{eq:volteqnpert})). In the first stage, the load time constants and the stable system  state matrix are estimated using WLS regression and thereafter the scaled Jacobian matrix is calculated. From the scaled Jacobian matrix, the initial estimates for the line parameters ($G_{ij}^{n,p*}$ and $B_{ij}^{n,p*})$ are calculated using WLS regression. The second stage involves using the Broyden method explained in section (\ref{Broyden}) to improve the initial estimates  $G_{ij}^{n,p}$ and $B_{ij}^{n,p}$ The proposed method in this paper is further summarized in Algorithm $1$\color{black}. 
        \begin{algorithm}
        \caption{The Two-Stage Real-Time Line Parameter Estimation for Multi-Phase Unbalanced Distribution Network} 
        \begin{algorithmic}[1]
            % \color{blue}

        \topPhase{Estimate initial parameters}
        % \REQUIRE 
\begin{flushleft}        \textbf{Input:}$P_{i}^{n}$,$P_{j}^{n}$,$Q_{i}^{n}$,$Q_{j}^{n}$,$V_{i}^{n}$,$V_{j}^{n}$,$\delta_{i}^{n}$,$\delta_{j}^{n}$,  $\forall n\in{\{a,b,c\}}$ obtained or calculated from $\mu$-PMUs 
        
        % \ENSURE  
       \textbf{Output:}$[{G}_{ij}^{n,p*}],[{B}_{ij}^{n,p*}]$ $\forall n,p\in{\{a,b,c\}}$
\end{flushleft}
        \\ \textbf{define} N= number of branches, {$\forall$ branch $\{i, j\} \in \mathscr{S}$} %\color{red} need to change notation \color{black}\color{blue}
        \\ \textbf{compute} ${A}$ using (\ref{eq:Amatrix}) -  (\ref{samplemean})

        \For {k =1:N}
        \State Estimate $\tau_{ppi}^{n}$ ,$\tau_{qqi}^{n}$,$\tau_{ppj}^{n}$,$\tau_{qqj}^{n}$ using (\ref{eq:tauest})
        \State  Estimate Initial ${G}_{ij}^{*,n,p}$,${B}_{ij}^{*,n,p}$ using (\ref{eq:GBest}) %\color{red} I think you need to estimate all initial G, B for all branches before going to stage 2? \color{black}
        \EndFor
        %next stage

        \phase{Estimate final parameters}
        % \REQUIRE 
       \begin{flushleft}   \textbf{Input:}$P_{i}^{n},P_{j}^{n},Q_{i}^{n},Q_{j}^{n},V_{i}^{n},V_{j}^{n}, [{G}_{ij}^{n,p*}],[{B}_{ij}^{n,p*}] \forall {n,p}\in{\{a,b,c\}}$ obtained or calculated from $\mu$-PMUs 
        % \ENSURE  
        
        \textbf{Output:}$[{G}_{ij}^{n,p}],[{B}_{ij}^{n,p}]$ $\forall n,p\in{\{a,b,c\}}$
        \end{flushleft}  
        \State Compute ${G}_{ij}^{n,p},{B}_{ij}^{n,p}$ using (\ref{eq:fine_est})-(\ref{eq:updateG})\\
        \Return ${G}_{ij}^{n,p},{B}_{ij}^{n,p}$ 
        \end{algorithmic} 
        \end{algorithm}
        \color{black}

\vspace{-0.15cm}
%\color{red} Given the space we have, I will keep only Algorithm 1, The framework can be moved to the journal version. But in Algorithm 1, please label stage 1 and stage 2. Also, delete the event detection. \color{black}

%\subsection{Summary of the Proposed Estimation Framework}
% We posit a two-step estimation process as shown in figure (\ref{fig:framework}); where, an initial estimation of the line parameters is conducted using the Ornstein-Uhlenbeck regression process and the results serve as a fine initial start for the quasi-Newton method (Broyden) employed in %analytically 
% obtaining more accurate results. 
 % In the first step, a representative dynamic load model is transposed to a multivariate Ornstein-Uhlenbeck process and the resulting equations are solved using regression to initially (step $I$) obtain a value for  $[G_{ij}$ and  $[B_{ij}]$. Next, Broyden method is used to fine-tune the estimation results in step $II$. In step $III$, event detection is implemented to detect when changes in the line parameters exceed a predetermined threshold value.

%\begin{figure}[1ht]
%\centering
%\includegraphics[width=0.7\columnwidth]{frame_work_oct.png}
%\caption{Framework for the Implementation of the Proposed Estimation Algorithm}  %\color{red} make it smaller with higher resolution \color{black}}
%\label{fig:framework}
%\end{figure}

\subsection{Performance Evaluation}
We evaluate the accuracy of line parameter estimates using the mean absolute percentage error (MAPE). %To compare our approach with state-of-the-art methods, 
We assess %the performance of the proposed and existing methods by comparing 
the MAPE %and estimation time $t(s)$ 
between estimated and true values across all branches in the distribution network. This evaluation is conducted through Monte Carlo simulations under varying noise levels. For example, $\text{MAPE}(\mathbf{G}_{ij}) = \frac{100\%}{N} \sum_{i=1}^{N} \frac{\left| \mathbf{G}_{ij}^\text{true} - \mathbf{G}_{ij}^\text{estimated} \right|}{\left| \mathbf{G}_{ij}^\text{true} \right|}$, where $N$ is the number of branches estimated.
% We use the mean absolute percentage error (MAPE) to evaluate the accuracy of the line parameter estimates. Performance comparison with state of the art is %\color{blue}
% stochastically evaluated by comparing the MAPE  and estimation time $t(s)$ between  the estimated and the true values over all the branches in the distribution network using Montecarlo simulations with different noise levels. %\color{red}estimated value and the true value over .... Monte Carlo\color{black}. 
% For example, %proposed method and the least absolute shrinkage and selection operator (Lasso) and adaptive Lasso methods in \cite{8601410}.
% \small{
%  \begin{equation}
%      \begin{aligned}
%          \text{MAPE}(\mathbf{G}_{ij}) = \frac{100\%}{N} \sum_{i=1}^{N} \frac{\left| \mathbf{G}_{ij}^\text{true} - \mathbf{G}_{ij}^\text{estimated} \right|}{\left| \mathbf{G}_{ij}^\text{true} \right|} 
%      \end{aligned}
%  \end{equation}
%  \normalsize}
%  where $N$ is the number of branches estimated.
 
\section{Results and Discussion}\label{RandD}

\subsection{Test System}
\begin{figure}[ht]
\includegraphics[width=8.5cm]{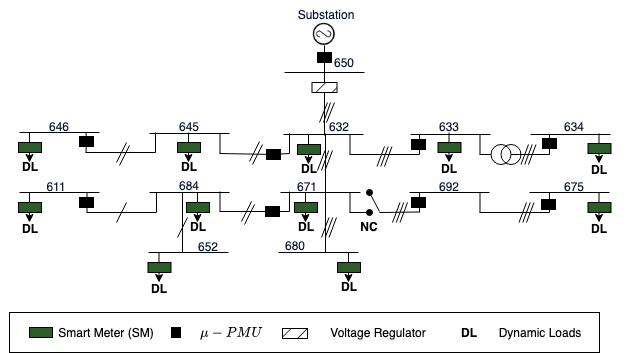}
\caption{IEEE $13$-bus Test Feeder (Number of slashes = Number of phases connecting two nodes.)}
\label{benchmark}
\end{figure}

The proposed algorithm is tested on the benchmark IEEE 13-bus multi-phase unbalanced distribution system (see Fig. (\ref{benchmark}) ), modeled in OpenDSS \cite{OpenDSS}. Operating at a nominal voltage of 4.16 kV, this %multiphase unbalanced 
network includes a switch to simulate topology changes affecting the admittance matrix. Real load data is obtained from the ADRES dataset \cite{Ardakanian}, which provides $1 s$ measurements of real and reactive power for 30 Austrian households over 14 days. Real-life PV data from \cite{PV30minutes}  is interpolated to match this $1 s$ load data. Three-phase loads are split into single-phase loads and randomly connected to all buses except the source and switch nodes. A three-phase 800 kVA PV system with a unity power factor is connected to bus 680. %Smart meters at each bus measure active and reactive power, with
$\mu$-PMU measurements at each bus are simulated using OpenDSS-MATLAB.

%The proposed algorithm is tested on an IEEE benchmark distribution test bus-13  in Fig. (\ref{benchmark}) 
% \cite{916993}
%. The multiphase unbalanced IEEE-13 is modelled in OpenDSS \cite{OpenDSS}. This system operates at a nominal voltage of $4.16 kV$ and contains a switch that can be operated to simulate a topology change leading to a modification in the admittance matrix. Real data from the ADRES dataset  \cite{Ardakanian} was used to model loads %\color{red} you mean load? \color{black} 
%connected to the feeders. This dataset contains high resolution $1 s$ measurements  of real and reactive power of $30 $ Austrian households over $14$ days. %\color{red} Does the dataset also contains PV data? Otherwise, cite the PV data source\color{black}
%The PV data is available at \cite{PV30minutes} with a granularity of $30$ minutes but is interpolated to $1s$ data to match the load data. The available three-phase loads are split into three single-phase loads and a random number of loads are connected to all the buses except the source bus and the switch node. A three-phase PV rated at $800 kVA$ with a unity power factor, is connected to bus $680$. Smart meters measuring active and reactive powers are assumed to be present at every bus and $\mu$-PMU measurements are simulated using loadflow in OpenDSS-MATLAB.

% \begin{figure}[1ht]
% \includegraphics[width=8.5cm]{Benchmark_IEEE13.png}

\subsection{Case Studies}
Two distinct case studies were conducted to evaluate Algorithm I's response under different grid conditions:
    \begin{enumerate}
        \item Case I: The original network configuration served as the baseline (Base Case).
        \item Case II: The effect of incorporating distributed energy resources (DERs) was investigated.
    \end{enumerate}
In the first case, $N = 3600$ samples of $V$ and $\delta$ are generated by running AC power flow simulations in OpenDSS-MATLAB. Gaussian noise with zero mean and variance $\sigma^{2}$ is added to the active and reactive power loads 
$P$ and $Q$. Following the power flow simulations, node voltages are treated as 
$\mu$PMU measurements, with additional Gaussian noise introduced to simulate measurement error. This noise adheres to IEEE Standard C37.118-2005, which specifies that PMU measurement errors should remain below 1\% of total vector errors (TVE).

%In the first case, $N = 3600$ samples of $V$ and $\delta$ are generated from running AC power flow in OpenDSS-MATLAB. Gaussian noises of zero mean and variance $\sigma^{2}$ are added to the active and reactive power loads $P$ and $Q$. After the power flow simulations, we treat the node voltages as $\mu$PMU measurements and once more introduce white Gaussian noise with zero mean and variance  $\sigma^{2}$ to imitate the measurement error. The noise is added in accordance with IEEE Standard C37.118-2005  \cite{4483699} which states that PMU measurement errors shall be lower than $1\%$ of total vector errors (TVE). 

\begin{figure}[!h]
    \centering
    \includegraphics[width=0.7\linewidth]{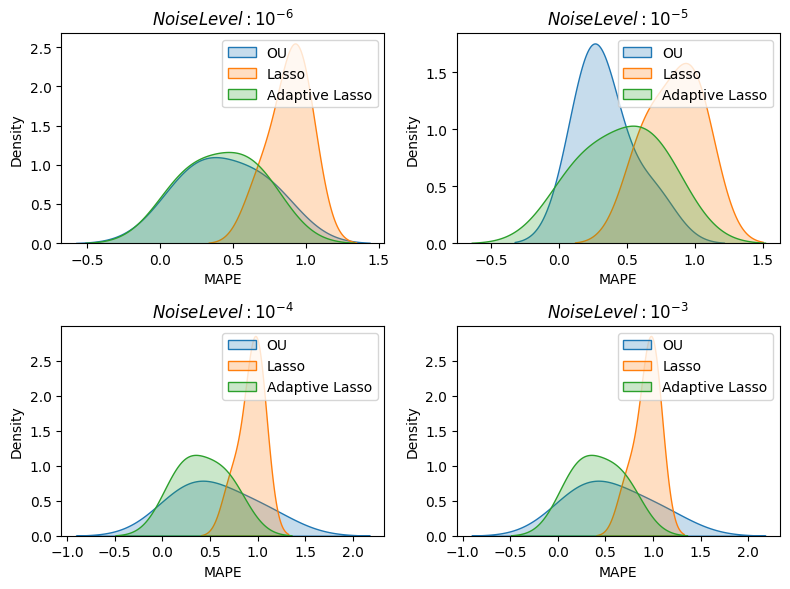}
    
    \caption{Estimation Error for Conductance: Case I}
    \label{fig:case1-mape_g}
\end{figure}

\begin{figure}[!h]
    \centering
    \includegraphics[width=0.7\linewidth]{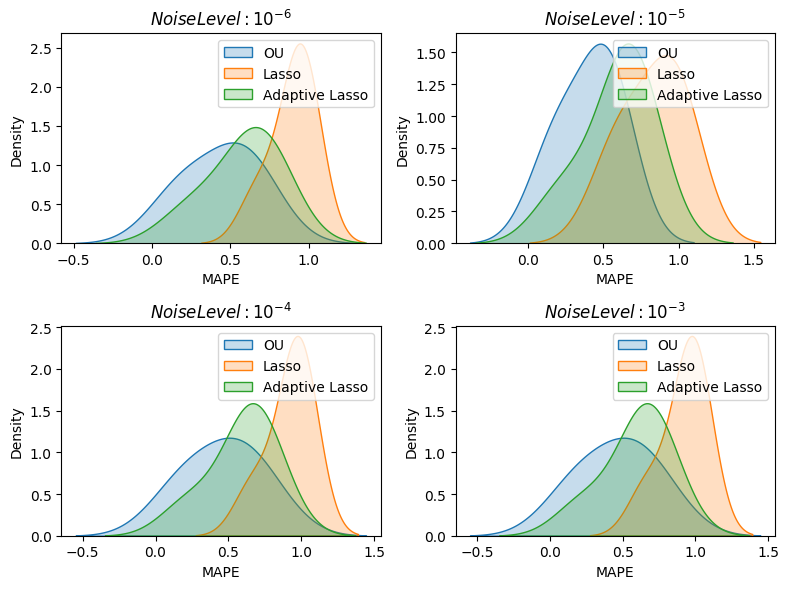}
    \caption{Estimation Error for Susceptance: Case I}
    \label{fig:case1-mape_b}
\end{figure}

The performance of the proposed algorithm is evaluated
statistically. %using a Monte Carlo method because of the randomness of the added noise. Hence, 
Particularly, 100 Monte Carlo simulation runs are conducted for each of the measurement noise levels, %several values of 
$\sigma =\{10^{-6},10^{-5},10^{-4},10^{-3}\}$. %\color{red} cite a paper to justify the intensity chosen\color{black} 
%\color{red} As shown in Fig. ...., what are presented under different measurement noise level.(briefly explain the compared metrics) (Also, did you run Monte-Carlo? You should briefly explain this)....\color{black} 
Figure. (\ref{fig:case1-mape_g}) and (\ref{fig:case1-mape_b}) present the average MAPE of line conductances and susceptances across all branches from the 100 Monte Carlo simulations 
at different noise levels. %Line parameters are estimated by branch and by figures (\ref{fig:case1-mape_g}) and (\ref{fig:case1-mape_b}).  %\color{red}correct it\color{black}
%color{red}The lower of the 
 The lower MAPE mean and narrower distribution for the proposed method indicate better performance. Compared to the proposed method, the density for the Lasso \cite{8273895} is concentrated at higher MAPE values and this becomes more pronounced at higher noise levels. While the adaptive Lasso \cite{8601410} shows slightly higher MAPE values, its performance remains comparable to the proposed method especially at lower noise levels.
\color{black}
% It can be observed that the proposed method outperforms the Lasso %\color{blue}
% \cite{8273895} %\color{red}cite\color{black} 
% and adaptive Lasso %\color{red}cite\color{black} method 
% \cite{8601410} in estimating the line susceptances and give comparable or better performance in estimating the line conductances, especially at low noise levels. % at lower noise levels 
%and is comparable with the adaptive lasso method at higher noise levels. \color{blue} The three algorithms show better performance for the estimation of susceptances; however, with increasing noise levels, there is a shift to the right in the pdf plots of Lasso and adaptive Lasso indicating worsening performance i.e higher errors with increasing noise levels  . 
Additionally, the proposed method is approximately 100 times faster than adaptive Lasso, as shown in Table (\ref{tab:comparison}), making it advantageous for online monitoring applications and in calculating real-time control actions on the grid. 

\begin{figure}[!h]
    \centering
      \includegraphics[width=0.7\linewidth]{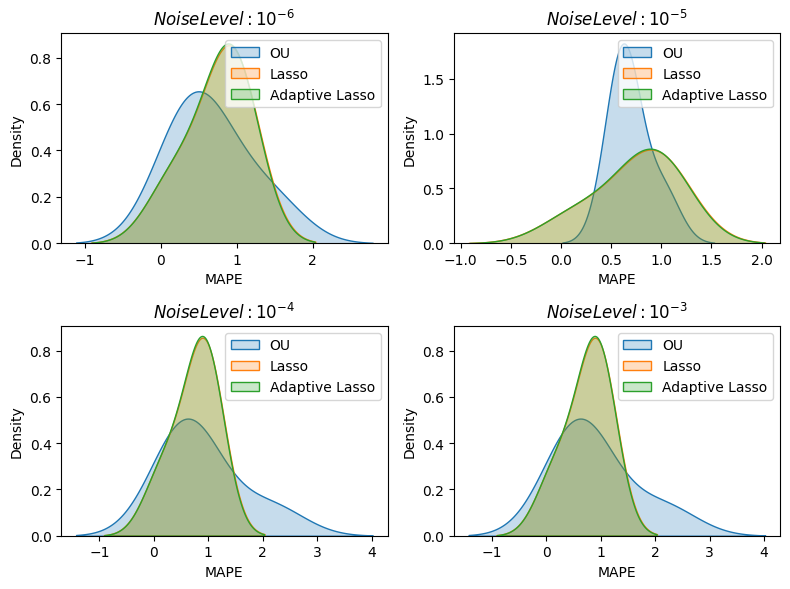}
    \caption{Estimation Error for Conductance: Case II}
    \label{fig:case2-mape_g}
\end{figure}

\begin{figure}[!h]
    \centering
        \includegraphics[width=0.7\linewidth]{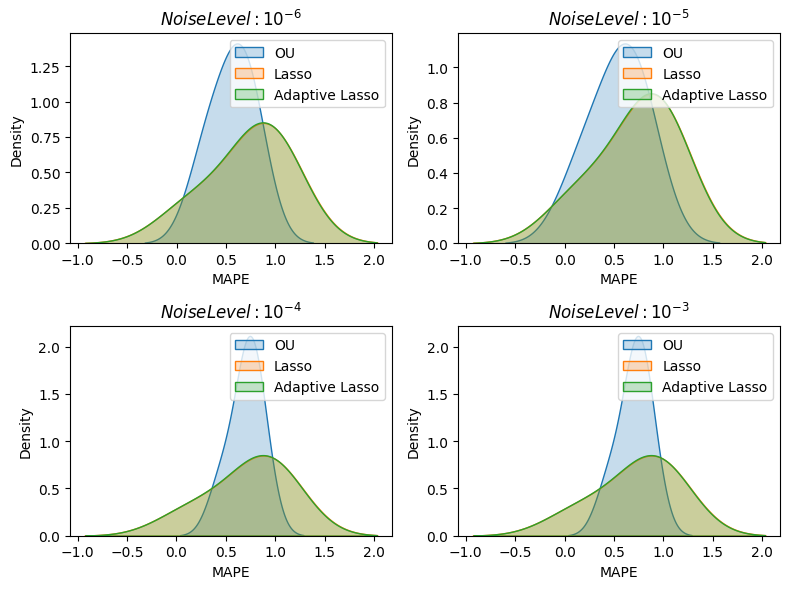}
    \caption{Estimation Error for Susceptance: Case II}
    \label{fig:case2-mape_b}
\end{figure}

In the second case, we test the estimation in the presence of volatile PV generation, using real-world PV data from \cite{PV30minutes} with added measurement noise, as in Case I. The results in Figures (\ref{fig:case2-mape_g}) and (\ref{fig:case2-mape_b})  show similar trends to those observed in Case I. Notably, with volatile PV, adaptive Lasso performs almost identically to Lasso, offering minimal improvement. In contrast, the proposed method consistently provides more accurate and more consistent estimation results in most cases. Moreover, it requires two orders of magnitude less computational time than adaptive Lasso, making it significantly more efficient. However, the 13-bus network offers a limited scope to test the proposed methodology; future work will investigate the scalability  to distribution systems of greater size and complexity.
% The scalability of this method to larger distribution systems is a topic for our future works. 

\color{black}

% \begin{table}[!h]
%     \centering
%     \captionsetup{justification=centering} 
%     \caption{
%     %COMPARISON OF 
%     COMPUTATIONAL TIME (s) %BETWEEN THE  PROPOSED METHOD AND \\
%     %THE METHODS IN \cite{8601410} 
%     - CASE I}
%     \begin{tabular}{c|c|c|c|c}
%     \hline
%     \diagbox{\textbf{Method}}{\textbf{Noise Level}}& $10^-{6}$ & $10^-{5}$ & $10^-{4}$ & $10^-{3}$ \\
%     \hline
%     OU & \textbf{1.65} & \textbf{1.39} & \textbf{1.24} & \textbf{1.72}  \\
%     \hline
%     Lasso & 59.50 & 51.60 & 51.03 & 48.69  \\
%     \hline
%     Adaptive Lasso & 167.54 & 155.48 & 162.47 & 164.53   \\
%     \hline
%     \end{tabular}\label{table:time_base}
%     \end{table}

\begin{table}[h] 
\centering
\caption{Computational Time(s) - Case I}
\begin{tabular}{lcccc}
\toprule
\textbf{Noise Level} & $10^-6$ & $10^-5$ & $10^-4$& $10^-3$ \\
\midrule
OU & $\mathbf{1.65}$ & $\mathbf{1.39}$ & $\mathbf{1.24}$ &  $\mathbf{1.72}$ \\
Lasso & $59.50$ & $51.60$ & $51.03$ & $48.69$ \\
Adaptive Lasso & $167.54$ & $155.48$ & $162.47$  & $164.53$ \\
\bottomrule
\end{tabular}
\label{tab:comparison} % Optional: Add a label for referencing
\end{table}

\vspace{-0.15cm}

% \begin{table}[!h]
% \centering
% \captionsetup{justification=centering} 
% \caption{
% %COMPARISON OF 
% COMPUTATIONAL TIME (s) %BETWEEN THE PROPOSED METHOD AND\\ 
% %THE METHODS IN \cite{8601410}
% - CASE II }
% \begin{tabular}{c|c|c|c|c}
% \hline
% \diagbox{\textbf{Method}}{\textbf{Noise Level}} & $10^-{6}$ & $10^-{5}$ & $10^-{4}$ & $10^-{3}$ \\
% \hline
% OU & \textbf{3.86} & \textbf{2.78} & \textbf{2.12} & \textbf{3.65} \\
% \hline
% Lasso & 104.74 & 151.88 & 195.15 & 92.97 \\
% \hline
% Adaptive Lasso & 260.45 & 389.26 & 475.44 & 736.70 \\
% \hline
% \end{tabular}\label{table:time_PV}
% \end{table}

\begin{table}[h] 
    \centering
    \caption{Computational Time(s) - Case II}
    \begin{tabular}{lcccc}
    \toprule
    \textbf{Noise Level} & $10^-6$ & $10^-5$ & $10^-4$& $10^-3$ \\
    \midrule
    OU & $\mathbf{3.86}$ & $\mathbf{2.78}$ & $\mathbf{2.12}$ &  $\mathbf{3.65}$ \\
    Lasso & $104.74$ & $151.88$ & $195.15$ & $92.97$ \\
    Adaptive Lasso & $260.45$ & $389.26$ & $475.44$  & $736.70$ \\
    \bottomrule
    \end{tabular}
    \label{tab:comparison2} % Optional: Add a label for referencing
    \end{table}

\section{Conclusion}
%\color{red} update conclusion \color{black}
In this paper, a two-stage line estimation method
for multiphase unbalanced distribution has been proposed. %line parameter estimation algorithm based on the multivariate Ornstein Uhlenbeck regression theorem was proposed. \color{blue}
Simulation results using real-life load and PV data demonstrate that the proposed method  can provide accurate estimation for line susceptances and conductances while reducing computational time by one to two orders of magnitude compared to existing methods. %compared to the two state of the art algorithms showed better performance with lower estimation errors at lower noise levels and comparable performance at higher noise levels in the presence of dynamics and uncertainties of renewable energy sources. Additionally, the computation time of the proposed method is $\frac{1}{100^{th}}$ the time of the compared algorithms making it more advantageous for online applications. \color{black}
Further work includes the application of the proposed method in monitoring and controlling larger distribution systems.
%\color{red} reduce refs. \color{black}

\bibliography{refs}
\bibliographystyle{ieeetr}

\vspace{12pt}

\end{document}